\documentclass[letterpaper, amsfonts, superscriptaddress, amssymb, amsmath, preprint, nofootinbib]{revtex4-1}
\pdfoutput=1

\usepackage{graphicx}
\usepackage[dvipsnames]{xcolor}
\usepackage[version=4]{mhchem}

\begin{document}

\title{Crystal Net Catalog of Model Flat Band Materials}

\author{Paul M. Neves}
\affiliation{Department of Physics, Massachusetts Institute of Technology, Cambridge, Massachusetts 02139, USA}
\author{Joshua P. Wakefield}
\affiliation{Department of Physics, Massachusetts Institute of Technology, Cambridge, Massachusetts 02139, USA}
\author{Shiang Fang}
\affiliation{Department of Physics, Massachusetts Institute of Technology, Cambridge, Massachusetts 02139, USA}
\author{Haimi Nguyen\footnote{Current Affiliation: Department of Chemistry, Columbia University}}
\affiliation{Department of Physics, Mount Holyoke College, South Hadley, Massachusetts 01075, USA}
\author{Linda Ye\footnote{Current Affiliation: Department of Applied Physics, Stanford University, California 94305, USA}}
\affiliation{Department of Physics, Massachusetts Institute of Technology, Cambridge, Massachusetts 02139, USA}
\author{Joseph G. Checkelsky}
\email{checkelsky@mit.edu}
\affiliation{Department of Physics, Massachusetts Institute of Technology, Cambridge, Massachusetts 02139, USA}

\date{March 4, 2023}

\maketitle

\section*{Abstract}
Flat band systems are currently under intense investigation in quantum materials, optical lattices, and metamaterials. These efforts are motivated by potential realization of  strongly correlated phenomena enabled by frustration-induced flat band dispersions; identification of candidate platforms plays an important role in these efforts.  Here, we develop a high-throughput materials search for bulk crystalline flat bands by automated construction of uniform-hopping near-neighbor tight binding models. We show that this approach captures many of the essential features relevant to identifying flat band lattice motifs in candidate materials in a computationally inexpensive manner. We apply this algorithm to 139,367 materials in the Materials Project database and identify 63,076 materials that host at least one flat band elemental sublattice. We further categorize these candidate systems into at least 31,635 unique flat band crystal nets and identify candidates of interest from both lattice and band structure perspectives. This work expands the number of known flat band lattices that exist in physically realizable crystal structures and classifies the majority of these systems by the underlying lattice, providing new insights for familiar ($\textit{e.g.}$ kagome, pyrochlore, Lieb, and dice) as well as previously unknown motifs.

\section{Introduction}
Flat band systems have recently received significant attention as platforms to realize novel quantum states. Theoretically, the non-dispersive nature of these infinitely massive flat bands may enable electronic correlation effects including ferromagnetism, high temperature fractional quantum Hall physics, topological superconductivity, and excitonic insulating behavior \cite{micnas1990superconductivity, mielke1991ferromagnetic, sethi2021flat, aoki2020theoretical, tang2011high, bergholtz2013topological}. The field of flat band physics has been recently invigorated by the experimental identification of flat electronic bands in 2D moiré heterostructures \cite{cao2018unconventional, cao2018correlated, park2021flavour, li2021imaging, zhang2020flat, balents2020superconductivity}, bulk quantum materials \cite{kang2020topological}, circuit QED systems \cite{kollar2020line}, optical lattices \cite{taie2015coherent}, and photonic crystals \cite{leykam2018artificial}. Understanding the potential breadth of flat band platforms is thus a topic of significant interest.

Flat band-hosting crystal lattices were first proposed over thirty years ago, exemplified by models for the kagome, Lieb, pyrochlore, and dice lattices \cite{lieb1989two, mielke1991ferromagnetic, sutherland1986localization, bergman2008band, derzhko2015strongly, leykam2018artificial} and the Penrose tiling \cite{arai1988strictly}. Despite the apparent simplifications—\textit{e.g.}, being of nearest neighbor, single orbital and isotropic nature—taken by these theoretical models, their relevance in describing electronic structures of real materials is supported by a growing number of experimental studies \cite{li2018realization, yin2019negative, kang2020dirac, kang2020topological, liu2020orbital, meier2020flat, li2021dirac, han2021evidence, ye2021flat}. This suggests that a simple tight-binding approach can provide key guidance for identifying additional flat band systems, even in the presence of other atomic species, spin-orbit coupling, orbital degrees of freedom, and interaction effects not taken into account at the tight-binding level. More recent theoretical efforts have expanded flat band models to more exotic lattices such as the diamond-octagon \cite{pal2018nontrivial} and the Creutz \cite{mondaini2018pairing, kuno2020extended, he2021flat}, and introduced general models by which novel flat band lattices can be systematically generated such as via the line or split graph constructions \cite{miyahara2005flat, ma2020spin, ogata2021methods, cualuguaru2022general, chiu2022line}. In contrast to these broadened efforts, experimental realization in crystalline systems has been relatively scarce and has focused on the kagome prototype (\textit{e.g.} references \cite{li2018realization, yin2019negative, kang2020dirac, kang2020topological, liu2020orbital, meier2020flat, li2021dirac, han2021evidence, ye2021flat}).  There is therefore an opportunity to expand flat band studies with the identification of candidates for other lattice motifs, especially those that can be found in realistic material structures.

Here, we develop a high-throughput approach to identify flat band systems by building simple (\textit{i.e.} nearest-neighbor, single orbital, uniform hopping) tight binding models on candidates drawn from the Materials Project (MP).  Motivated by the experimental observations of elementally derived flat bands in recent flat band studies \cite{kang2020dirac, kang2020topological}, we then identify each elemental sublattice that hosts a non-trivial flat band originating from destructive interference of compact localized eigenstates (CLS) \footnote{Here a trivial flat band is defined as one which is caused by a component of the lattice which is isolated from other components and does not contain any hopping paths to infinite distance.} \cite{jain2013commentary}. As the bandstructure of a uniform hopping model relies only upon the connectivity of the underlying periodic graph (the ``crystal net"), we are able to categorize the majority of these systems by employing a crystal net isomorphism testing algorithm (Systre) \cite{delgado2003identification} that generates a unique canonical key for each distinct crystal net. The remaining are grouped by invariant quantities of the sublattice graph and bandstructure (see methods section B and supplemental note II). This allows us to identify the common flat band lattice motifs present in the MP, and to identify certain elements, spacegroups, and chemical structures that host candidates of a given flat band lattice, providing a broad new set of flat band lattices for theoretical and experimental study.

\section{Results}
\subsection{Search Overview}
\begin{figure*}
	\includegraphics[width=\textwidth]{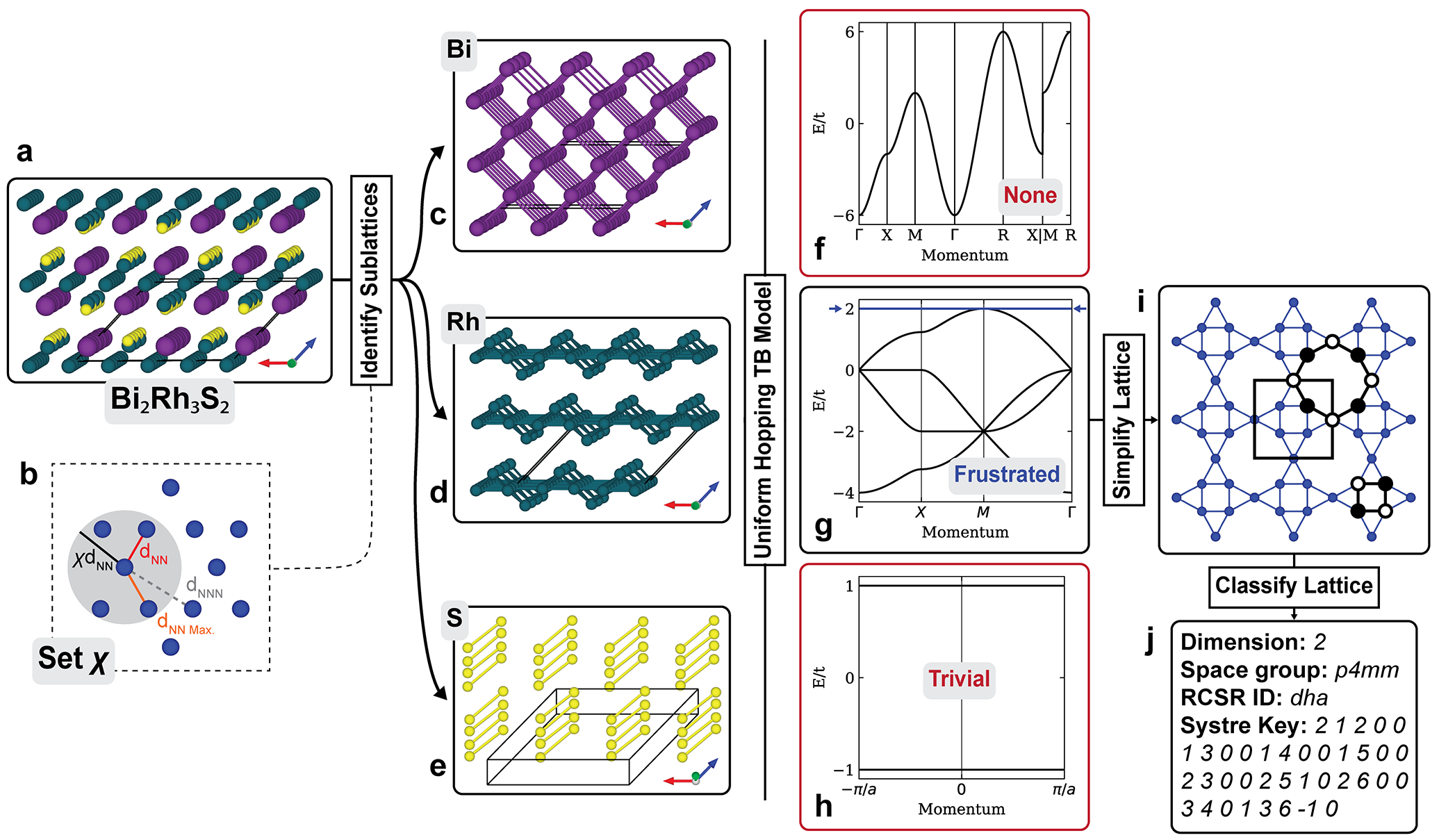}
	\caption{\label{fig:flowchart} \textbf{Flowchart of the tight binding flat band search algorithm.} \textbf{a} Structure of candidate \ce{Bi2Rh3S2} (spacegroup $C2/m$, lattice parameters $a$=11.49 Å, $b$=8.42 Å, $c$=8.13 Å, $\alpha=90^\circ$, $\beta=133.45^\circ$, and $\gamma=90^\circ$). \textbf{b} Nearest neighbor (NN) distance $d_{NN}$, the bond length cutoff $\chi d_{NN}$, the longest distance included as a bond $d_{NN, \textrm{ max}}$, and the shortest distance excluded as a next nearest neighbor (NNN) bond $d_{NNN}$. \textbf{c-h} The Bi lattice of \ce{Bi2Rh3S2} supports no flat bands (\textbf{c,f}), while the S contains only trivially localized molecular flat bands (\textbf{e,h}). The Rh sublattice  contains a non-trivial or frustrated two-dimensional flat band lattice with a doubly degenerate flat band at energy $E=2t$ for hopping $t$ which supports two distinct CLSs, the square kagome (or ``squagome") lattice, identified as ``dha" in the reticular chemistry structure resource \cite{o2008reticular} (\textbf{d,g}). \textbf{i} The squagome represented in high symmetry form, and \textbf{j} identified based on its Systre classification key. The unit cell of the squagome is shown with a black outline, and the two unique CLSs are shown with black and white lattice sites, indicating alternating signs of the wavefunction on each site (all other sites being zero).}
\end{figure*}

The algorithm employed here is outlined in Fig. \ref{fig:flowchart}. For each material in the MP (\textit{e.g.}, \ce{Bi2Rh3S2} in Fig. \ref{fig:flowchart}a), we consider each individual elemental sublattice within the crystal structure (Fig. \ref{fig:flowchart}c-e). For the shortest nearest neighbor (NN) distance $d_{NN}$ between any two atoms of this species in this structure, we identify all pairs of atoms in this sublattice less than or equal to a multiple $\chi$ of $d_{NN}$ apart (depicted in Fig. \ref{fig:flowchart}b). Next, we build a tight binding model with uniform self-energy and one orbital at each atomic site in the sublattice with hopping energy $t>0$ between all site pairs $\langle i,j \rangle$ such that the site pair distance $d_{ij}<\chi d_{NN}$ (Fig. \ref{fig:flowchart}f-h):
\begin{equation}
    H = -t \sum_{\langle i,j\rangle} \left(c_i^\dag c_j + c_j^\dag c_i\right).
\end{equation}
$c_i$ ($c^\dagger_i$) is the fermion creation (annihilation) operator at site $i$. We further define $d_{NNN}$ as the shortest atomic distance greater than $\chi d_{NN}$ (the shortest bond not included in the tight binding model).

A search of 468,378 individual elemental sublattices from 139,367 materials in the MP for near-neighbor isotropic uniform hopping tight binding flat bands was performed. The calculation was performed for $\chi$ cutoffs of 1.02, 1.05, 1.1, 1.2, and 1.4 (the statistics reported throughout this work include results of all listed choices of $\chi$). 63,076 unique material entries were found to contain at least one non-trivially localized flatband across the entire 3D brillouin zone in at least one of their elemental sublattices for at least one value of $\chi$. Some materials contain multiple decoupled lattice components (a component is defined as a connected subgraph that is not part of any larger connected subgraph) or multiple flatband lattices among different elemental sublattices in the same material; in total 108,341 flat band models were found within elemental sublattices. 15,288 unique flat band crystal nets were identified with Systre \cite{delgado2003identification}, while 68,710 components evaded classification in Systre due to barycentric node collisions (see methods). Applying lattice invariant based classification schemes identifies at least 16,347 additional crystal net groups, yielding at least 31,635 unique flat band crystal nets.

\subsection{Observed Flat Band Motifs and Abundances}
In Table \ref{tab:table1} we list in descending order the abundances of materials hosting the 10 most common flat band lattices (the most common lattices and their tight binding bandstructure are shown in Fig. \ref{fig:common_lats} and Fig. \ref{fig:lat_bs}, respectively, with additional common lattices shown in the supplementary Fig. S2).  The kagome lattice is the most commonly identified lattice, found in 5,329 materials (Fig. \ref{fig:common_lats}a and \ref{fig:lat_bs}a). Second is the pyrochlore lattice with 3,700 materials (Fig. \ref{fig:common_lats}b and \ref{fig:lat_bs}b). The one dimensional stub lattice is third with 3,669 (Fig. \ref{fig:common_lats}c and \ref{fig:lat_bs}c), followed by the ``diamond chain" \cite{aoki2020theoretical} lattice, a 1D chain of alternating one or two sites, with 2,557 (Fig. \ref{fig:common_lats}d and \ref{fig:lat_bs}d). Following this are the Lieb lattice with 1,976 (Fig. \ref{fig:common_lats}e and \ref{fig:lat_bs}e), the fluorite \ce{CaF2} lattice (with both the Ca and F sites occupied by the same species) with 1,330 (Fig. \ref{fig:common_lats}f and \ref{fig:lat_bs}f), a diamond chain lattice with a transverse bond here referred to as the ``Kite" lattice with 1,291 (Fig. \ref{fig:common_lats}g and \ref{fig:lat_bs}g), the tetrahedron chain with 1,131 (Fig. \ref{fig:common_lats}h and \ref{fig:lat_bs}h), a honeycomb with an additional orbital connected to one vertex, which we refer to here as the ``Stub Honeycomb" with 1,049 (Fig. \ref{fig:common_lats}i and \ref{fig:lat_bs}i), and the hyperkagome \cite{lawler2008gapless} lattice which is the edge net of the \ce{SrSi2} lattice \cite{friedrichs2003regular} has 930 (Fig. \ref{fig:common_lats}j and \ref{fig:lat_bs}j).

In Table \ref{tab:table2} we provide the abundances of elemental flatband sublattices being hosted by \textit{s}, \textit{p}, \textit{d}, or \textit{f} block elements on 1-, 2-, or 3- dimensional networks.  In terms of composition, most identified flat band lattices are built of \textit{p} block elements, followed by the \textit{d} block, \textit{s} block, and \textit{f} block (see also supplementary Fig. S1). We hypothesize that this is likely biased by the contents of the database itself (for example, 50.6\% of materials in the MP contain oxygen at the time of writing); it is of interest to explore the role of chemical bonding in this (\textit{i.e.} if covalent bonds may be more likely to form the relatively non-close-packed flat band lattices depicted in Fig. \ref{fig:common_lats} than a metallic element). We also observe a relatively uniform distribution of one-dimensional, two-dimensional, and three-dimensional flat band lattices. Only 20.8\% of results contain a flat band with the most restrictive choice $\chi = 1.02$.

\subsection{Identifying Flat Band Material Families}
\begin{figure*}
	\includegraphics[width=\textwidth]{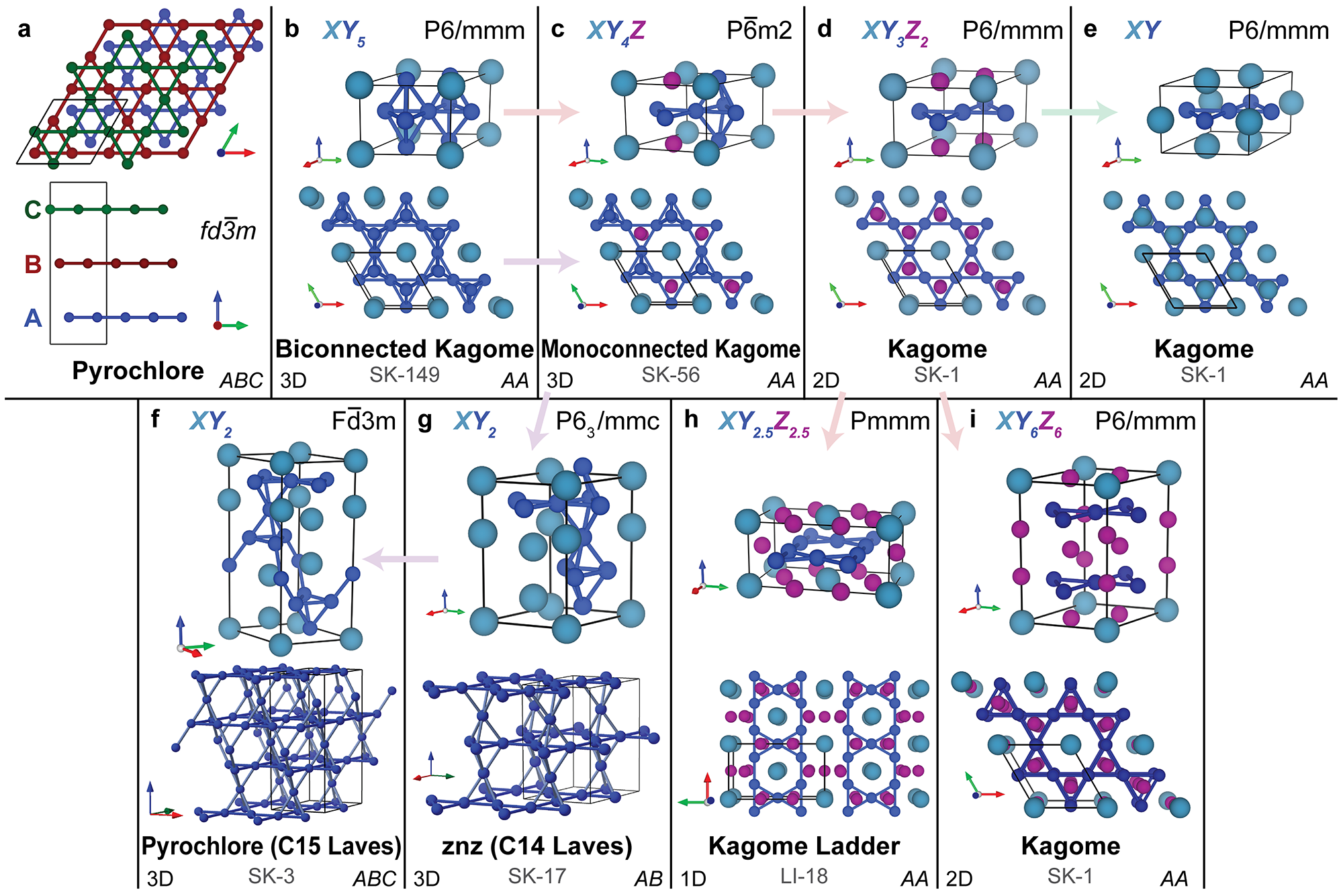}
	\caption{\label{fig:relations} \textbf{Relations between flat band lattices in the kagome families.} \textbf{a} The pyrochlore \textit{ABC} stacking of kagome lattices (interlayer sites not shown). \textbf{b} The \ce{CaCu5} type structure, showing the 3D flat band \textit{Y} sublattice. \textbf{c} By substituting one \textit{Y} site from (b), a different 3D flat band lattice is obtained. \textbf{d} Substituting a second \textit{Y} site creates the 2D kagome flat band lattice. \textbf{h} Substituting further to \textit{XY$_{2.5}$Z$_{2.5}$} creates a 1D version of the kagome lattice, referred to here as the ``Kagome ladder." \textbf{e} The \textit{XY} kagome flat band lattice can be generated from \textbf{d} by shifting the corner \textit{X} site and converting \textit{Z} sites to \textit{X} sites. \textbf{i} The \textit{XY$_6$Z$_6$} lattice is obtained from \textbf{d} by doubling the unit cell along the c-axis and splitting one \textit{X} site into two \textit{Z} sites. \textbf{g} The znz (C14 hexagonal Laves phase) 3D flat band lattice is created from \textbf{c} by changing to an \textit{AB} stacking, substituting an \textit{X} for a \textit{Z}, and distorting two \textit{X} sites. \textbf{f} By altering to an \textit{ABC} stacking, the cubic C15 Laves pyrochlore lattice can be realized in the \textit{XY$_2$} system. The pyrochlore unit cell depicted in \textbf{a} and \textbf{f} is rotated such that the $c$-axis is along the conventional [111] direction to facilitate comparison amongst structures. Dimensionality is indicated in the lower left of each panel. Stacking order is indicated in the lower right of each panel. The lattice code of each lattice is indicated in the lower center of each panel.}
\end{figure*}

Analysis of the identified flat band lattice materials reveals connections between several of these networks in terms of dimensionality and the character of the CLS. For example, many flat band networks arise from a similar interference mechanism as the kagome net: a ring with an even number of sites with each nearest neighbor of the ring hopping to two adjacent sites within the ring. In Fig. \ref{fig:relations}, we show the relationship between various 1D, 2D, and 3D flat band lattice materials that fit this description. Starting with the \textit{XY$_5$} (\ce{CaCu5}-type) structure (Fig. \ref{fig:relations}b) of \textit{AA} stacked bi-connected kagome layers, replacement of one or both connecting \textit{Y} atoms with \textit{Z} atoms yields the mono-connected (\textit{XY$_4$Z}, Fig. \ref{fig:relations}c) and disconnected (\textit{XY$_3$Z$_2$}, Fig. \ref{fig:relations}d) kagome flat band lattices, respectively. Further substitution of \textit{Y} for \textit{Z} to \textit{XY$_{2.5}$Z$_{2.5}$} results in a 1D kagome ladder flat band lattice (Fig. \ref{fig:relations}h). Converting \textit{Z} to \textit{X} in \textit{XY$_4$Z} to obtain \textit{XY$_2$}, different stackings (with slight distortion on \textit{X} sites) produce the the \textit{ABC} stacked (\ce{MgCu2}-type, C15) cubic laves phase pyrochlore flat band lattice (Fig. \ref{fig:relations}f) and the \textit{AB} stacked (\ce{MgZn2}-type, C14) ``znz" hexagonal laves phase flat band lattice (Fig. \ref{fig:relations}g). Shifting, substituting, and splitting sites in the \textit{XY$_3$Z$_2$} structure can result in the \textit{XY} (CoSn-type) kagome flat band lattice (Fig. \ref{fig:relations}e), or the \textit{XY$_6$Z$_6$} (\ce{MgFe6Ge6}-type) kagome flat band lattice (Fig. \ref{fig:relations}i). This can be considered a kagome ``family'' of systems: all can be thought of as distinct ways of connecting kagome lattices while maintaining destructive interference of the hopping around the hexagonal ring. As such, the CLS's of all lattices in Fig. \ref{fig:relations} will consist of orbitals with alternating signs around the hexagonal ring. Further exploration of other motifs cataloged in the present search may provide insights into other flat band material families and how flat band changes across those classes.

\section{Discussion}
\begin{figure*}
	\includegraphics[width=\textwidth]{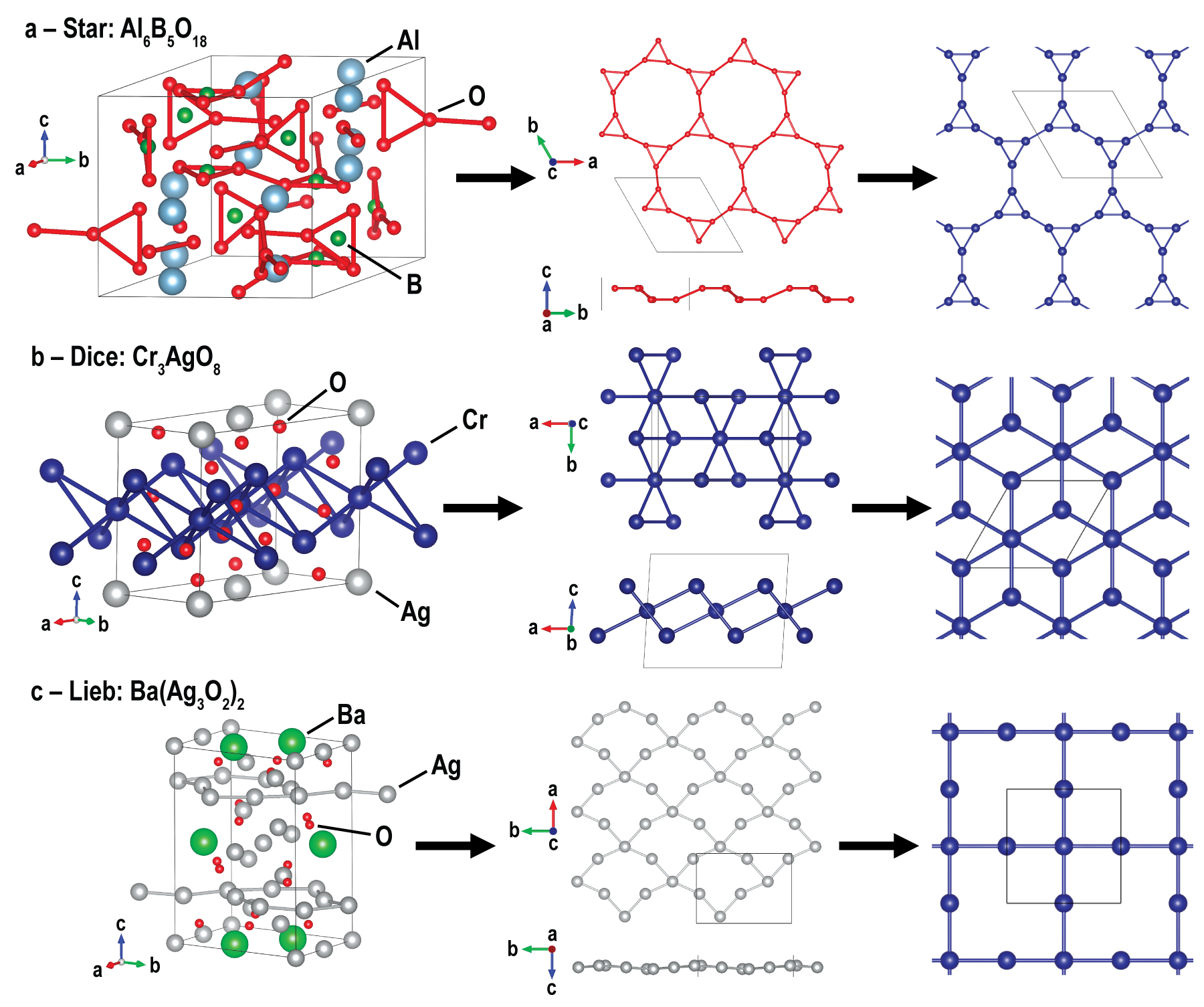}
	\caption{\label{fig:distorted} \textbf{Distorted lattices and higher symmetry flat band lattices.} \textbf{a} (left) Structure of \ce{Al6B5O18}, (center) its isolated oxygen star lattice, and (right) the topologically equivalent higher symmetry counter part. \textbf{b} Structure of \ce{Cr3AgO8}, its isolated chromium dice lattice, and undistorted form. \textbf{c} Structure of \ce{Ba(Ag3O2)2}, its isolated silver Lieb lattice, and undistorted form.}
\end{figure*}

We have identified 108,161 candidate sublattice flat band systems in the MP, comprised of more than 31,635 unique  lattices. These include motifs with a variety of potentially interesting bandstructure features including isolated flat bands, linear and quadratic intersections with the flat band (including singularity of the wavefunction at intersection points \cite{rhim2019classification}), and multiply degenerate flat bands. These lattices are categorized to the extent possible with current graph theory algorithms, allowing the identification of isomorphically equivalent flat band lattices in a majority of search results. The present graph theory-based approach enables further flexibility in the present search through its ability to identify distorted lattices and categorize them by their most symmetric forms. For example, as shown in Fig. \ref{fig:distorted}a, \ce{Al6B5O18} is identified to contain a hidden distorted oxygen star lattice, the chromium sublattice of \ce{Cr3AgO8} can be considered a 3D distorted dice lattice (Fig. \ref{fig:distorted}b), and the \ce{Ba(Ag3O2)2} unit cell contains two distorted Lieb lattices of silver (Fig. \ref{fig:distorted}c).  This provides a significantly more general framework for classifying lattices than a purely geometric method.

Beyond this classification, the present search also aids in identifying materials in different proximity to ideal models. The ratio $d_{NNN}/d_{NN,\textrm{ max}}$ is one potential metric of the robustness of an embedded flat band sublattice  (larger being more favorable). A tight grouping of bond lengths in a sublattice (with a large distance to the next nearest bond not considered) may likely represent a relatively undistorted or ideal lattice. Conversely, a model where $d_{NNN}$ is only slightly larger than $d_{NN,\textrm{ max}}$ is more likely to be highly distorted and/or overlook relevant hopping pathways. The shortest bond included, $d_{NN}$, should also be small (typically 3 Å) to ensure $t$ is a dominant energy scale. The robustness of a flat band to the inclusion of hopping decay is another metric of the flat band model reliability. One might also seek simple materials with a minimal number of sites in the unit cell to minimize additional bands (examples of complex unit cells are provided in supplementary Fig. S3). Combining these criteria, we provide a curated list in the supplemental materials of 2,759 materials with filtered values of key parameters: materials with 16 or fewer atoms per unit cell, $d_{NN} \leq 3.7\ \text{\AA}$, $d_{NN,\textrm{ max}}/d_{NN} \leq 1.05$, $d_{NNN}/d_{NN, \textrm{ max}} \geq 1.35$, and a negative energy per atom in the MP calculation are included.

In terms of potential shortcomings of the present approach, in distilling each elemental sublattice to its minimalistic model based on the topology of the crystal net, many effects that could render the flat band dispersive are overlooked. For example, this method does not include higher orbital angular momentum orbitals. In some cases this simplification is valid, for instance the $d_{z^2}$ orbital model matches the $s$ orbital model in CoSn \cite{kang2020topological}, but in others orbitally-enabled flat bands may be overlooked such as the $\{p_x,p_y\}$ honeycomb model \cite{wu2007flat}. Beyond this, spin orbit coupling (SOC), further neighbor and anisotropic hopping, interaction effects, and orbitals on other sublattices are neglected (while orbitals on the considered sublattice are given equal self-energies). In the future, this flat band search method could be extended to produce additional flat band models and more accurate results by including these effects. More sophisticated search algorithms are also of significant interest; in particular, comparison with recent geometry- and DFT-based catalogs \cite{meschke2021search, jovanovic2022simple, regnault2022catalogue} may provide new insights for both theoretical models and identification of experimental materials targets.

Alternative methods for screening tight binding models for candidate flat band systems could also be considered. Starting from \textit{ab initio} density functional theory calculations for a material, one can manually construct an effective Hamiltonian with maximally localized Wannier functions \cite{marzari2012maximally} (though this is significantly more computationally intensive than the present approach). Other methods attempting to construct a tight binding model database \cite{garrity2021database} or generate tight binding models from \textit{ab initio} calculations can also be considered \cite{unke2021se, zhang2021equivariant, nakhaee2020machine, vitale2020automated}. Finally, we propose that this general framework of high-throughput tight-binding and graph isomorphism analysis can be applied to a broader range of materials searches, for example, in searching for systems with symmetry protected band nodes, nodal points, or massless dispersions. The materials insight provided by those and the present flat-band focused investigations may provide a theoretical and computational resource for identifying exotic phases in lattice systems and an experimental resource for selection of synthesis targets in both artificial lattice and material-based studies.

\section{Methods}
\subsection{Identification of nontrivial flat bands}
We eliminate components (subsets of atoms and bonds in the periodic graph that do not connect via any hoppings to other components of the graph) that are completely isolated within the lattice -- \textit{i.e.}, clusters of connected atoms that cannot hop infinitely far from any of the atoms in the cluster, as exemplified by the S sublattice in \ce{Bi2Rh3S2} (Fig. \ref{fig:flowchart}h). As such a completely localized component of a crystal net will have no momentum dependent eigenvalues, this step identifies the number of ``trivial" flat bands that will be present in the final bandstructure (those that would describe a lattice of decoupled molecular states). All remaining flat bands must then admit a description via the construction of nontrivial CLS due to frustrated hopping, \textit{e.g.} the Rh sublattice in \ce{Bi2Rh3S2} in Fig. \ref{fig:flowchart}g (see supplementary note I).

Discarding sites in the lattice determined to be trivially localized, we then calculate the eigenvalues of the remaining tight binding Hamiltonian on a 16$\times$16$\times$16 grid mesh of \textit{k} points in the Brillouin zone. We bin the eigenvalues with bins one thousandth of the total range of the bandstructure. A flat band is identified for any energy bin that contains an eigenvalue in that bin at every calculated \textit{k} point. Thus, a band must be no more dispersive than 0.1\% of the total dispersion of the bandstructure across the entire Brillouin zone. Such a search is presented here for $\chi=$1.02, 1.05, 1.1, 1.2, and 1.4 in order to capture lattices with a wide range of distortion and connectivity. $\chi=1.02$ was used as the lowest cutoff to allow for up to 2\% bond distortions to be considered identical, and to accommodate rounding errors in the relaxed structure ($\chi=1.4$ was used as the largest cutoff to avoid inclusion of bonds along the hypotenuse of three atoms in a right triangle).

\subsection{Lattice classification}
We attempt to categorize the identified lattices into groups of identical crystal nets (\textit{e.g.} all kagome lattices). As the band flatness in a tight binding model with uniform hopping originates from the topology of the periodic graph that describes the connectivity of orbitals, lattices identified by this search can be categorized by their crystal net isomorphisms. As such, a highly distorted lattice (Fig. \ref{fig:flowchart}d) will have identical bandstructure with its undistorted version (Fig. \ref{fig:flowchart}i), assuming the connectivity of the orbitals is the same. To classify lattices, we use the method of barycentric placement implemented by Systre \cite{delgado2003identification}, which associates a Systre ``key" unique to each periodic graph, the ideal symmetry group of the graph, and its dimensionality (Fig. \ref{fig:flowchart}j). The limitation of this method is that for approximately 40\% of flat band lattices, two or more vertices' barycentric locations are identical which prevents those lattices from being classified according to this scheme. For these, we group lattices based on lattice invariants including the number of atomic sites, the number and energies of the flat bands, and the approximate range of the energy eigenstates (see supplemental note II). Each unique flat band lattice type identified in this work is assigned a lattice ID number beginning with ``LI-'' or ``SK-'' for those classified by lattice invariant or systre key respectively. Following this prefix, a number is assigned based on the frequency of occurrence of the lattice in descending order.

\subsection{Code and databases used}
We obtain crystal structures from the MP \cite{jain2013commentary} (database version May 13, 2021). Structural analysis is performed with the pymatgen python library \cite{ong2013python, ong2015materials}. Tight binding models are calculated with the pythTB python library \cite{yusufaly2013tight}. Crystal net classification and analysis is performed with Systre \cite{delgado2003identification}. Additional analysis is performed with standard python libraries, Numpy, and Scipy.

\section{Acknowledgements}  We thank J. Cano, B.-J. Yang, T. Suzuki, G. Tritsaris, D. Tabor, and A. Aspuru-Guzik for useful discussions.  This research is funded in part by the Gordon and Betty Moore Foundation EPiQS Initiative, through Grant GBMF9070 to J.G.C. (computation), NSF grant DMR-2104964 (statistical analysis), and AFOSR grant FA9550-22-1-0432 (crystallographic analysis). H. N. and L.Y. acknowledge support by the STC Center for Integrated Quantum Materials, NSF grant number DMR-1231319.
This work was performed in part at the Aspen Center for Physics, which is supported by National Science Foundation grant PHY-1607611.

\section{Author Contributions}
P.M.N. wrote the code with input from J.P.W. and S. F.. P.M.N., J.P.W., and S.F. interpreted the search results with input from L.Y.. H.N., J.P.W., and L.Y. performed preliminary geometry-based searches. J.G.C. supervised the project. All authors analyzed the results and contributed to writing the manuscript.

\section{Competing Interests Statement}
The authors declare no competing interests.

\bibliography{FB_bibl}

\begin{figure}
	\includegraphics[width=\textwidth]{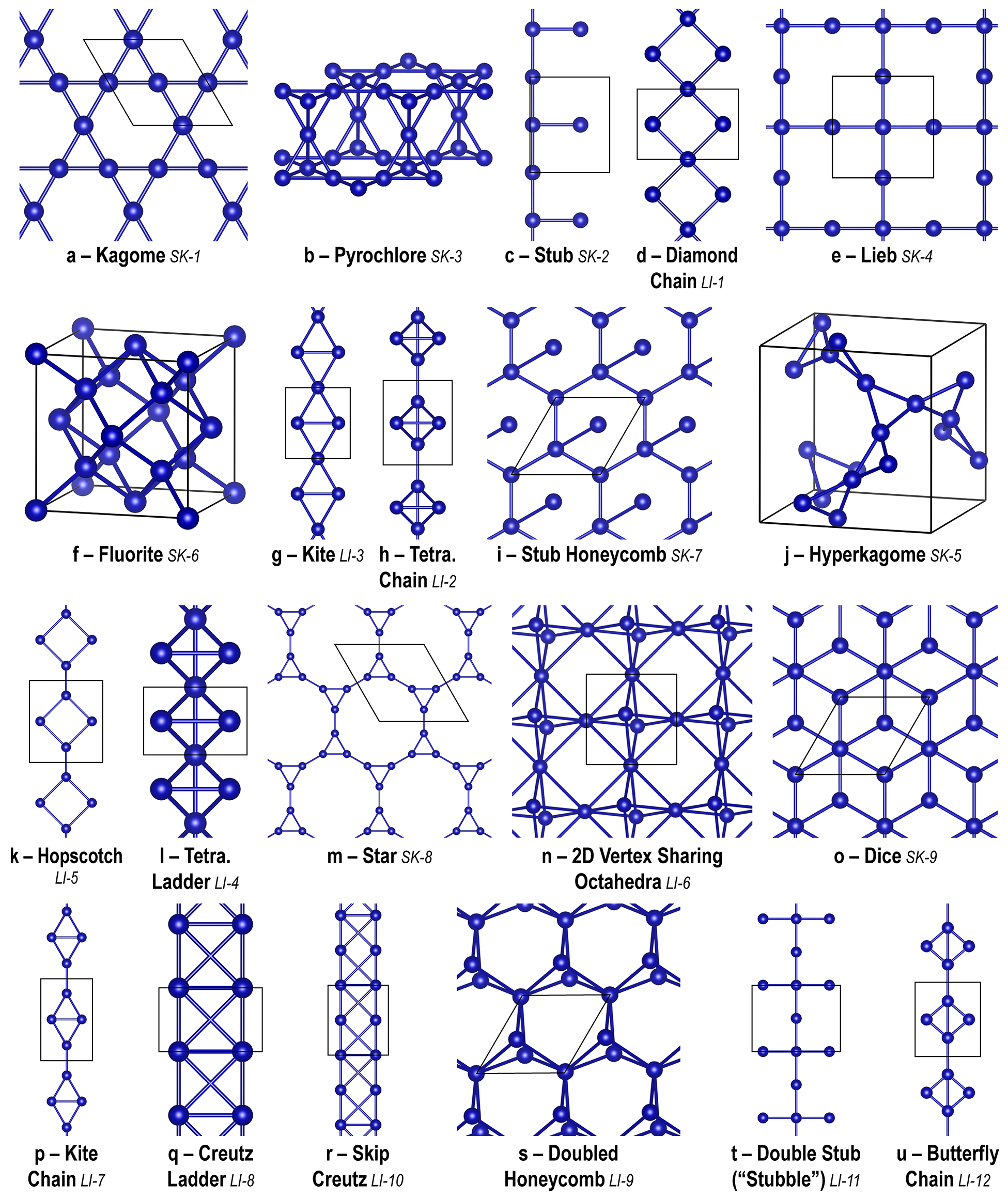}
	\caption{\label{fig:common_lats} \textbf{The twenty one most common non-trivial flat band lattices.} \textbf{a-u} Lattices identified by the present search sorted in descending order of occurence. The unit cell for each is shown with a solid black line.  The lattice code of each lattice is indicated after the lattice name.}
\end{figure}

\begin{figure*}
	\includegraphics[width=0.9\textwidth]{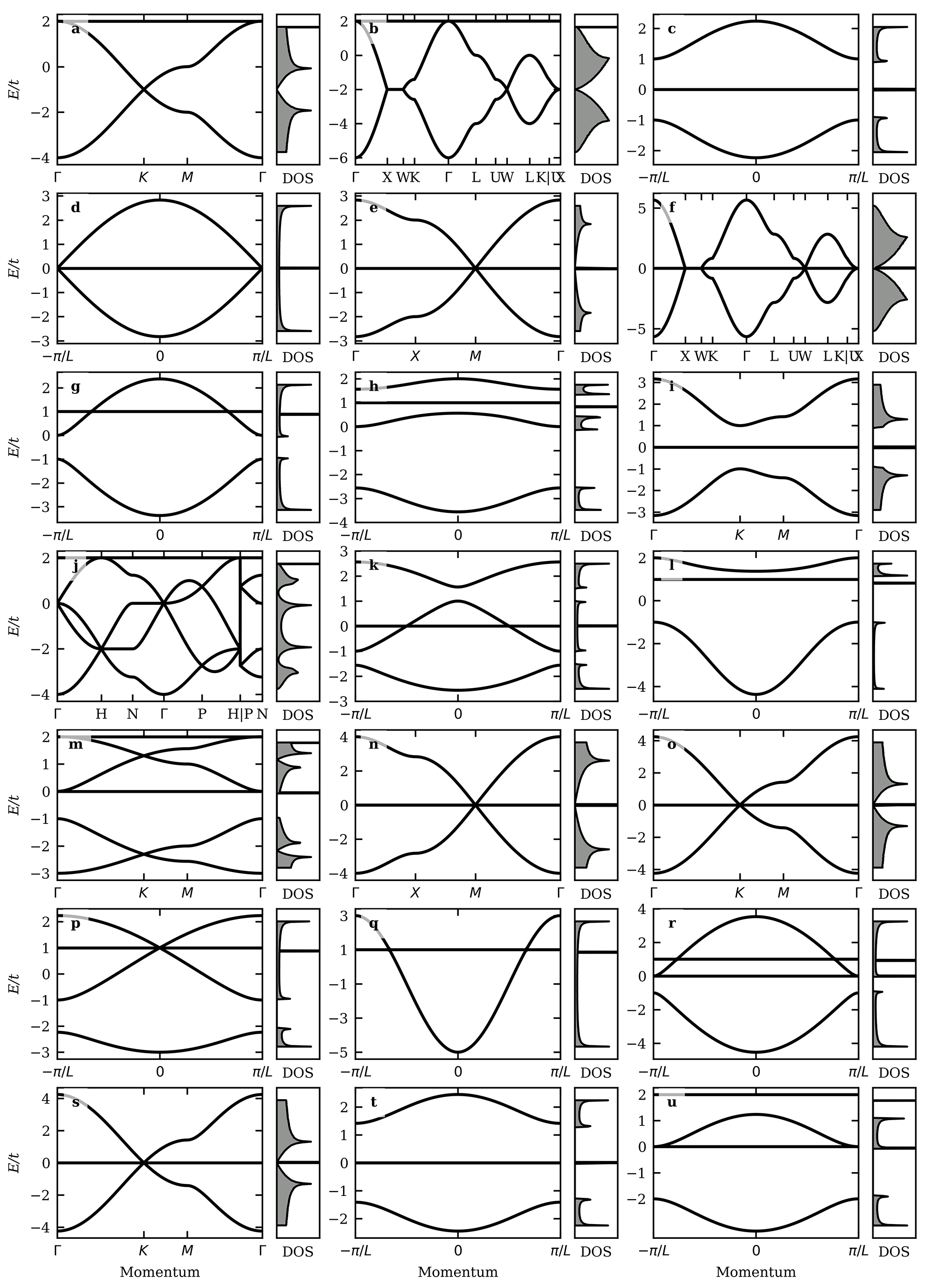}
	\caption{\label{fig:lat_bs} \textbf{Bandstructure and density of states for the most common lattices.} Bandstructures follow the same panel ordering \textbf{a} to \textbf{u} as presented in Fig. \ref{fig:common_lats}. The energy $E$ is normalized by the hopping integral $t$. The right hand sub-panels show the density of states (DOS).}
\end{figure*}

\begin{table}
\caption{\label{tab:table1} Statistics for the number of materials containing the most common flat band lattices.}
\begin{tabular}{cc}
\hline \hline
\textrm{Lattice} & \textrm{No. Materials}\\
\colrule
All & 139,367\\
Have F.B. & 63,076\\
Curated & 2,759\\
Kagome & 5,329\\
Pyrochlore & 3,700\\
Stub & 3,669\\
Diamond Chain & 2,557\\
Lieb & 1,976\\
Fluorite & 1,330\\
Kite & 1,291\\
Tetrahedron Chain & 1,131\\
Stub Honeycomb & 1,049\\
Hyperkagome & 930\\
\hline \hline
\end{tabular}
\end{table}

\begin{table*}
\caption{\label{tab:table2} Statistics for general properties of elemental sublattices identified to contain flat band models.}
\begin{tabular}{cc}
\hline \hline
\textrm{Property} & \textrm{No. Sublattices}\\
\colrule
All & 468,378\\
Have F.B. & 108,161\\
Curated & 2,958\\
\textit{s} block & 15,021\\
\textit{p} block & 65,050\\
\textit{d} block & 24,868\\
\textit{f} block & 3,222\\
1D & 35,537\\
2D & 36,692\\
3D & 36,558\\
Systre compatible & 60,913\\
Systre incompatible & 47,248\\
\hline \hline
\end{tabular}
\end{table*}

\end{document}

\title{Supplementary Information: \\ Crystal Net Catalog of Model Flat Band Materials}
\author{Paul M. Neves}
\affiliation{Department of Physics, Massachusetts Institute of Technology, Cambridge, Massachusetts 02139, USA}
\author{Joshua P. Wakefield}
\affiliation{Department of Physics, Massachusetts Institute of Technology, Cambridge, Massachusetts 02139, USA}
\author{Shiang Fang}
\affiliation{Department of Physics, Massachusetts Institute of Technology, Cambridge, Massachusetts 02139, USA}
\author{Haimi Nguyen\footnote{Current Affiliation: Department of Chemistry, Columbia University}}
\affiliation{Department of Physics, Mount Holyoke College, South Hadley, Massachusetts 01075, USA}
\author{Linda Ye\footnote{Current Affiliation: Department of Applied Physics, Stanford University, California 94305, USA}}
\affiliation{Department of Physics, Massachusetts Institute of Technology, Cambridge, Massachusetts 02139, USA}
\author{Joseph G. Checkelsky}
\email{checkelsky@mit.edu}
\affiliation{Department of Physics, Massachusetts Institute of Technology, Cambridge, Massachusetts 02139, USA}
\date{\today}
\maketitle

\section{Supplementary Note: Detection of trivial flat bands}
We detect trivial flat bands by first identifying sets of sites $\nu_i$ that only hop to other sites within the set. Starting with site $\nu_1$ in cell $(0,0,0)$, we add each site that connects directly to $\nu_1$, keeping track of the cell each neighbor is in. If any site is present in this list in more than one unit cell, then it is possible to travel infinitely far on this component, and the test terminates. If no duplicate sites in different cells are found, then all sites connected to the sites already visited are added to the list, and another duplication check is performed. This search iterates for 5 total hops, and if no duplicate sites in different unit cells are found, the component is categorized as trivially localized.

\section{Supplementary Note: Classification scheme via graph invariants for lattices not handled by Systre}
For each lattice, we calculate the degeneracy and energy of each flat band, the maximum energy in the bandstructure, the minimum energy in the bandstructure, the number of sites, the number of edges, the coordination sequence for each site, and the TD10 (defined in The Reticular Chemistry Structure Resource (RCSR) as the sum of the coordination sequence \cite{o2008reticular}). When more than one node is present in a lattice, TD10 is the arithmetic mean of the sum of the coordination sequence of each node. Energies are obtained by randomly sampling the Brillouin zone along high symmetry directions and at 4096 random points in the Brillouin zone. The calculated band extrema will in general differ from the true band extrema due to random sampling, and the flat bands are only found to within one part in a thousand of the total bandwidth due to energy binning. Therefore, comparisons of band extrema and flat band energies are made with a 1\% tolerance with respect to the total bandwidth. For each node, the coordination sequence is the set of $n^{\textrm{th}}$ order topological neighbors (\textit{i.e.}, the number of nodes in the periodic graph whose minimum distance from the original node is $n$) calculated here to $n=10$ \cite{delgado2003identification}. Many of these lattice invariants (flat band degeneracy, number of sites, number of edges, and the number of coordination sequences) are extensive under unit cell doubling (tripling, etc) and this search often finds doubled lattices. As such, invariants are reduced to the minimum possible under the assumption that if there is a greatest common denominator greater than one for a lattice, it is doubled. The equivalence of two lattices' sets of invariants is a necessary, but not sufficient condition for their isomorphic equivalence. We note that when we apply the lattice invariant sorting method to the graphs categorized with Systre, the lattice invariant sorting mis-categorizes (is unable to differentiate two different graphs) 0.8\% (638/75,805) of flat band components and identifies 98.9\% (15,122/15,288) of the different crystal net isomorphisms, demonstrating that this method reflects the true relative abundances of lattices to a significant degree.

\section{Supplementary Note: Histograms}
\begin{figure}
	\includegraphics[width=0.9\columnwidth]{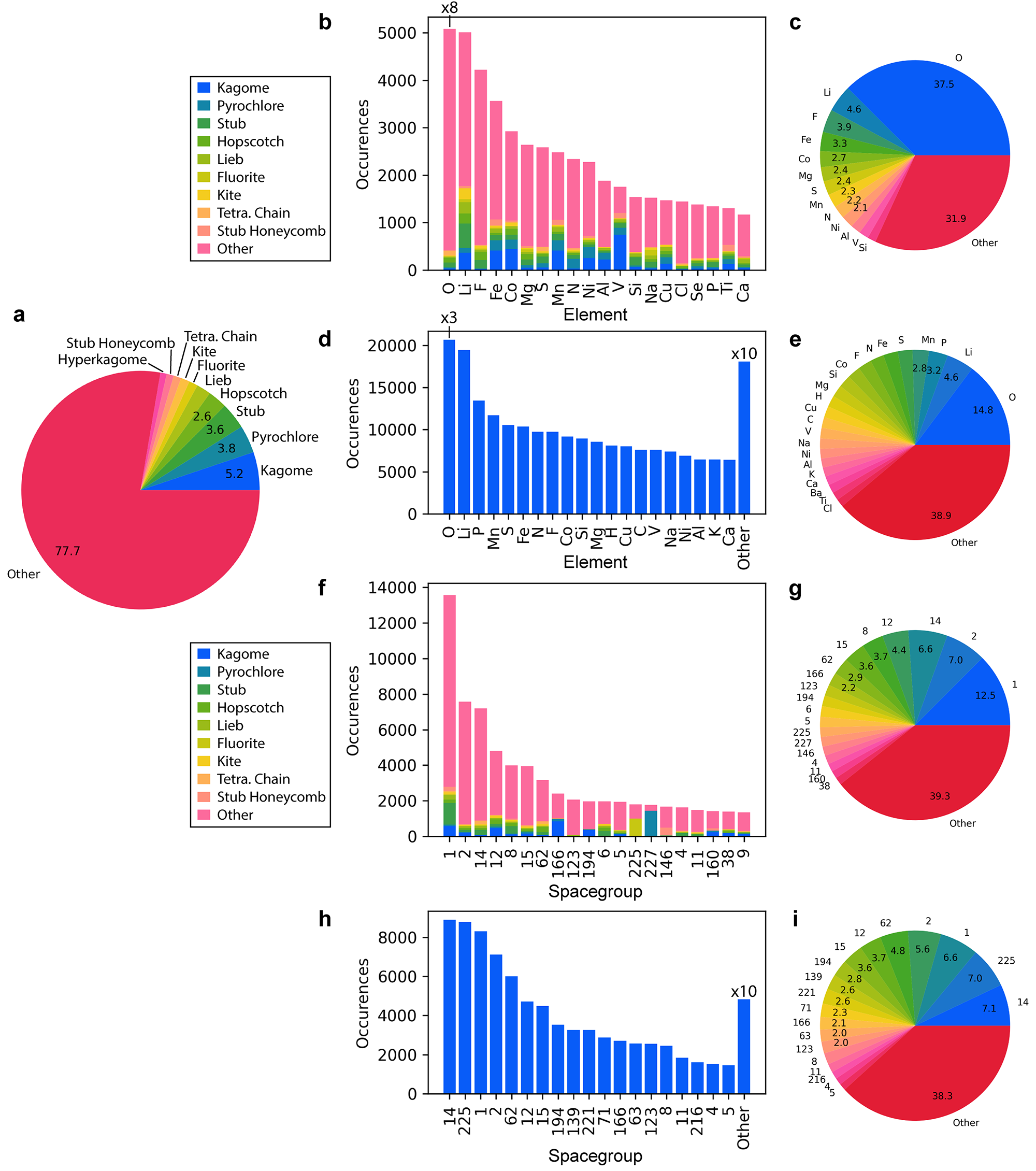}
	\caption{\label{fig:hists} Metaanalysis of search results. \textbf{a} Relative abundance of different flat band sublattices found in this search. \textbf{b} Bar and \textbf{c} pie charts of abundances of flat band sublattice elements, with different common flat band motifs shown in \textbf{b}. \textbf{d} Bar and \textbf{e} pie charts of the abundance of different elemental sublattices in the materials project. \textbf{f} Bar and \textbf{g} pie charts of abundances of flat band sublattice spacegroups, with different common flat band motifs shown in \textbf{f}. \textbf{h} Bar and \textbf{i} pie charts of the abundance of different space groups in the materials project.}
\end{figure}

We present histograms of different flat band lattices among different space groups and elements in Fig. \ref{fig:hists}. Oxygen sublattices tend to host the most flatband lattices (Fig. \ref{fig:hists}b-c), which is potentially explained by the abundance of oxygen in the MP (Fig. \ref{fig:hists}d-e) and the ability of oxygen to stabilize many complex structures. We note, however, that oxygen flat band sublattices must be examined with care because in many cases the oxygen acts as a ligand within a bipartite lattice in the material. Lithium is also a common flat band species (likely due to both the abundance of lithium materials in the MP and its reactivity). Many of the 3$d$ transition metals follow as recently studied in \textit{e.g.} transition metal Kagome compounds \cite{yin2019negative, kang2020dirac, kang2020topological, liu2020orbital, li2021dirac, han2021evidence, ye2021flat}. The triclinic spacegroup 1 ($P1$) is the most common in these results (Fig. \ref{fig:hists}f-g); we hypothesize that this is due to prevalence in the MP (Fig. \ref{fig:hists}h-i) and low symmetry. Certain spacegroups appear likely hosts for particular types of lattices: 1, 12, 166, and 194 host many Kagome lattices, 227 contains many Pyrochlores, 225 has many fluorites, 198 and 230 host the hyperkagome lattice, and 146 contains the stub honeycomb. We also see that Kagome and Pyrochlore lattices are common in 3$d$ transition metals.

\section{Supplementary Note: Additional Flat Band Lattices}
\begin{figure}
	\includegraphics[width=\columnwidth]{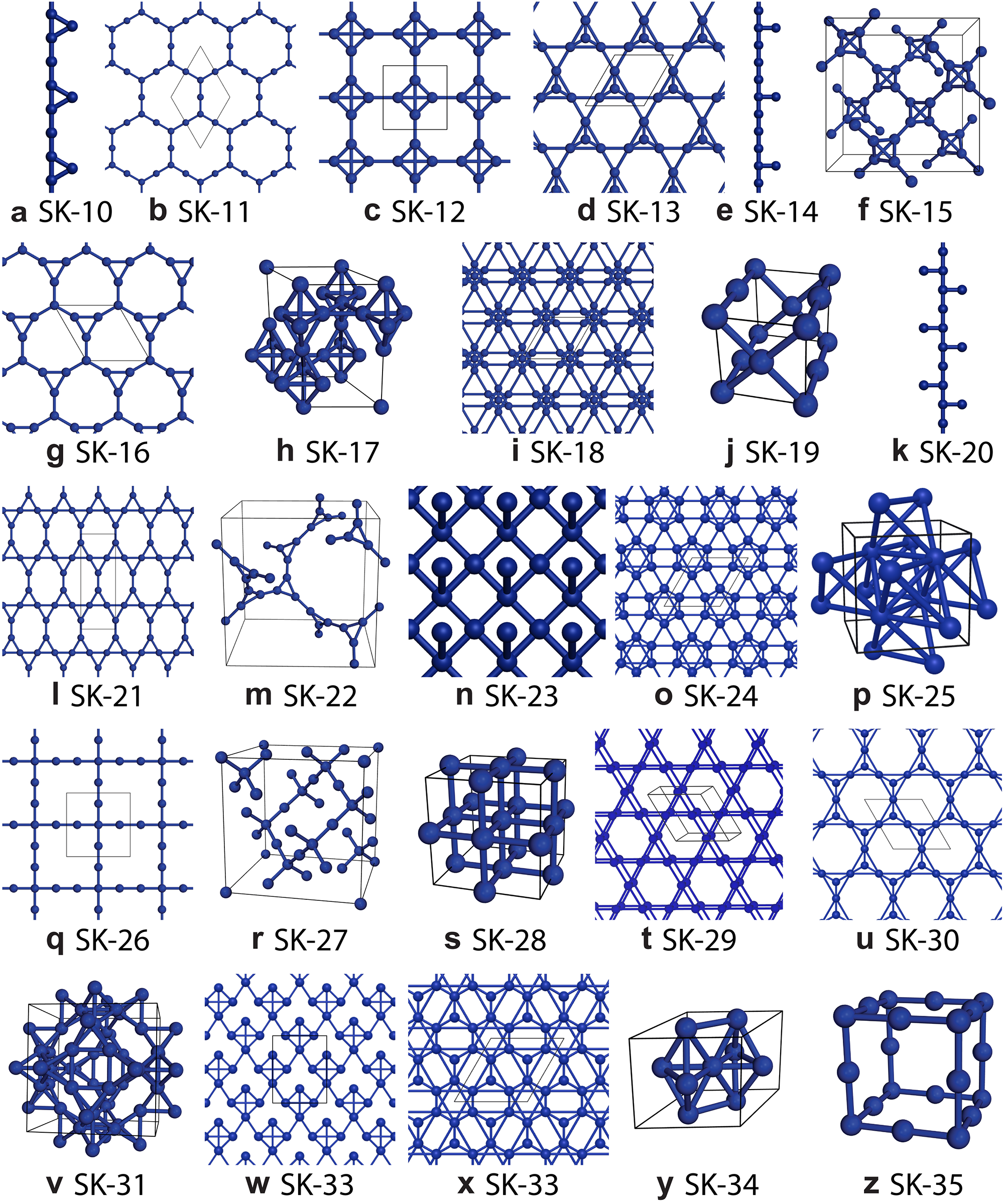}
	\caption{\label{fig:more_FBs} Additional flat band lattices. \textbf{a-z} Additional (systre-classified) flat band lattices found via this search, presented by abundance in descending order.}
\end{figure}

Additional flat band lattices uncovered in this search are shown in Fig. \ref{fig:more_FBs}. Many can be understood as originating from main text Fig. 2 with an additional bond, atom, or stub included. Fig. \ref{fig:more_FBs}h is the C14 Laves ``znz" lattice depicted in main text Fig. 4g. Fig. \ref{fig:more_FBs}m is the hyperkagome 2j with an additional bond between each triangle. Fig. \ref{fig:more_FBs}z is the 3D generalization of the Lieb lattice main text Fig. 2e, and Fig. \ref{fig:more_FBs}q is the Lieb with an additional atom in the chain between corners. Fig. \ref{fig:more_FBs}l is between the kagome (main text Fig. 2a) and star (main text Fig. 2m) lattices. Fig. \ref{fig:more_FBs}j is an AA stacking of Lieb lattices with one additional atom connecting the chain sites.

\section{Supplementary Note: Example of Complex Zeolite Flat Band Models}
\begin{figure}
	\includegraphics[width=\columnwidth]{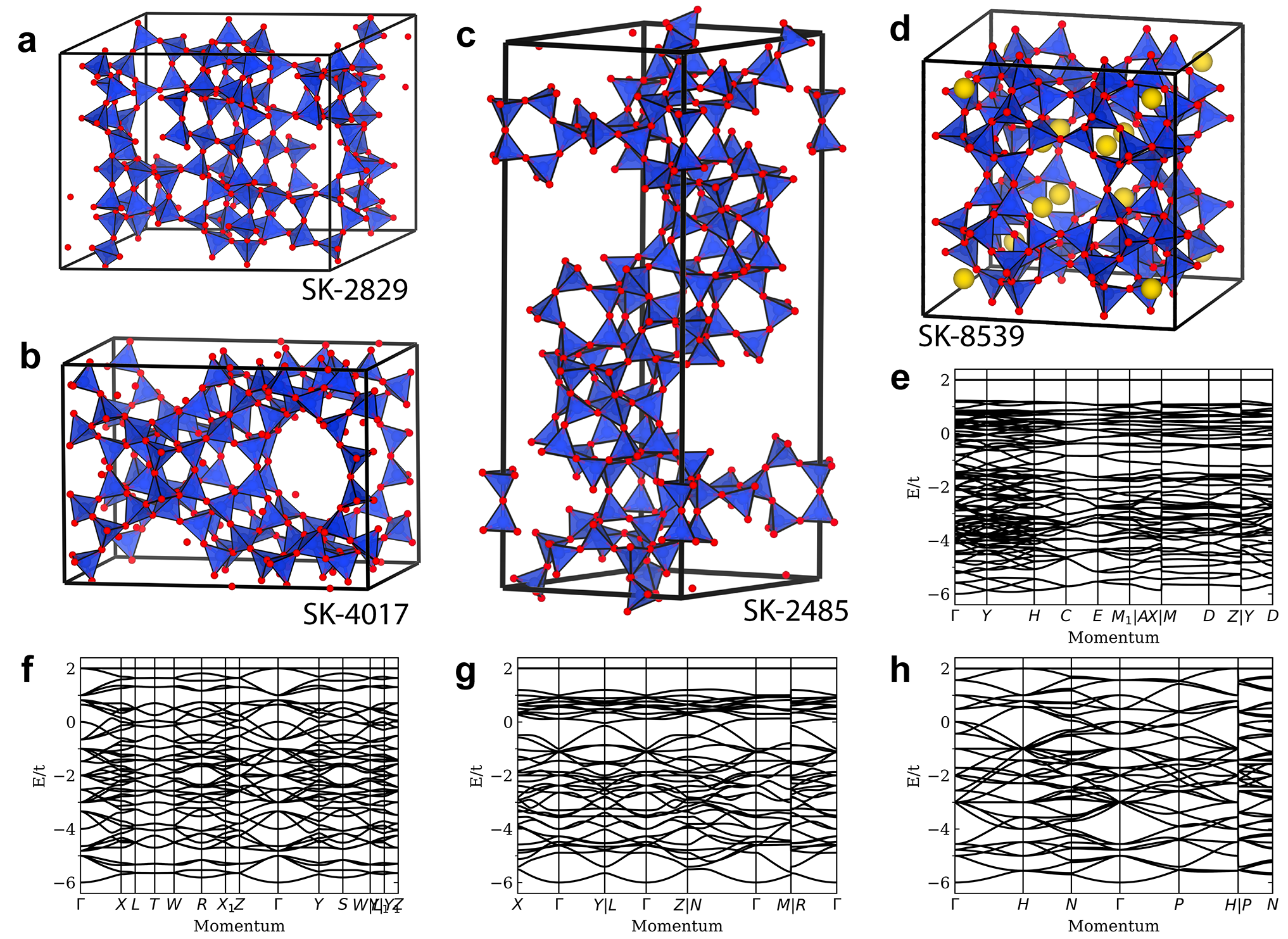}
	\caption{\label{fig:complex} Complex zeolite structures made of corner sharing tetrahedra. \textbf{a-d} Depiction of four complex zeolite structures. \textbf{e-h} Bandstructures corresponding to structures in \textbf{a-d}. An isotropic hopping nearest neighbor tight binding model constructed on the oxygen lattice with bonds between oxygen nearest neighbors (edges of the tetrahedra) contains flat bands due to frustrated hopping.}
\end{figure}

We show results of complex structures observed in Fig. \ref{fig:complex}. Each is a silicon dioxide based zeolite structure. The flat band tight binding model exists on the oxygen (red) sublattice, with bonds indicated by the edges of the blue tetrahedra surrounding each silicon atom. The lattice in Fig. \ref{fig:complex}a (\ce{SiO2} mp-1200292) contains 192 oxygen atoms per unit cell and 92 flat bands per unit cell in the oxygen NN model. With $d_{NNN}/d_{NN,max}>1.1$, the systems shown in Fig. \ref{fig:complex}b (\ce{Si48O107}, mp-1221354) and Fig. \ref{fig:complex}d (\ce{NaSi6O12}, mp-1212197)  have 48 flat bands each. The system in Fig. \ref{fig:complex}c (\ce{SiO2}, mp-1250639) has 360 atoms per unit cell (the largest of any result herein) and 120 out of the 240 bands are classified as flat.  We note that the flat bands in all of these models originate from the corner sharing tetrahedron nature of the structures, which results in frustrated hopping. As this model shares the same origin of frustration as the kagome and pyrochlore flat bands (loops of sites where any neighbor of the loop hops to two adjacent sites on the loop), the flat band energy is at $2t$ in all the zeolite structures in Fig. \ref{fig:complex}.

\bibliography{FB_bibl}

